\documentclass[12pt]{iopart}
\usepackage{graphicx}

\begin{document}

\title[Perfect and flexible quantum state transfer in the hybrid system atom coupled-cavity]{Perfect and flexible quantum state transfer in the hybrid system atom coupled-cavity}

\author{B. F. C. Yabu-uti$^1$, J. A. Roversi$^2$}

\address{Instituto de F\'isica "Gleb Wataghin", Universidade Estadual de
 Campinas, 13083-859, Campinas, SP, Brazil}
\eads{\mailto{$^1$yabuuti@ifi.unicamp.br}, \mailto{$^2$roversi@ifi.unicamp.br}}

\begin{abstract}

We investigate a system composed of $N$ coupled cavities and two-level atoms interacting one at a time. Adjusting appropriately the atom-field detuning, and make the hopping rate of photons between neighboring cavities, $A$, greater than the atom-field coupling $g$ (i.e. $A>>g$), we can eliminate the interaction of the atom with the nonresonant normal modes reducing the dynamics to the interaction of the atom with only a single-mode. As an application of this interaction, we analyze the transmission of an arbitrary atomic quantum state between distant coupled cavities. In the ideal case, we obtain a flexible and perfect quantum communication. Considering the influence of dissipation an interesting parity effect emerge and we obtain $N$ maximum in which it is still possible to achieve a quantum communication more efficient than a purely classical channel between the ends. We also studied important sources of imperfections in procedure execution.

\end{abstract}

\submitto{\jpb}

\maketitle

\section{\label{sec:level1}Introduction}
\

An important goal in the field of quantum information science, for distributing and processing quantum information, is the coupling of distant systems in order to achieve the transfer of quantum states (known or unknown) from one place to another.

Cavity quantum electrodynamics (CQED) provides a promising setting for processing quantum information  \cite{raimond,vahala,spillane,sauer,aoki,trupke,barclay}. In the strong coupling regime, that can be achieved between atoms and cavity field modes, the coherent part of the evolution dominates over the decoherence processes and quantum dynamics of the system can be observed. 
In fact much attention have been devoted to a system formed by $N$ (2 or more) cavities in an array close enough to allow photons hop to neighboring cavities due to the overlap of the local field modes or indirectly linked by an optical fiber \cite{skarja,serafini,angelakis,nohama,bose.angelakis,hartmann,yabuuti,ogden,irish,li}. 

Differently from schemes based on spin chains \cite{bose,osborne,burgarth,christandl,bose.contemp}, coupled cavities array (CCA) has the advantage of easily addressing individual lattices sites (for instance with optical lasers). Furthermore, atoms trapped inside (or close to the surface) cavities can have relatively long-lived atomic levels for encoding quantum information. Besides that, atoms and coupled cavities make a good model to investigate many-body phenomena as they offer several degrees of freedom to control the dynamics like the detuning, the atom-field coupling strength and the hopping rate \cite{angelakis,ogden,irish,zhang}.  

Two extreme situations can be considered in the hybrid system atom coupled-cavity: $A>>g$ and $g>>A$. In such limits, Zhang et. al. discuss the dynamics of one excitation in a system consisting of two two-level atoms each interacting with one of two coupled single-mode cavities \cite{zhang}.

For $g>>A$ is observed an effect of photon blockade in the array \cite{angelakis,irish}. This effect had recently lead to the prediction of a Mott insulator phase for polaritons (entangled states of atoms and photons) in CCA. In addition, the system simulates a XY spin model with the presence and absence of polaritons corresponding to spin up and down \cite{angelakis}. 
On the other hand, important works \cite{skarja,serafini,nohama,bose.angelakis,ogden,li} has been done on the problem of transferring atomic states between different coupled cavities . State transfer along an array of polaritons in a coupled-cavity system was studied in \cite{bose.angelakis}.

For $A>>g$, Nohama and Roversi \cite{nohama} investigated the transmission of a arbitrary quantum state working with two coupled cavities and two-level atoms in each one. They showed that for $\delta=-A$, where $\delta$ is the detuning between the atom and cavity field, a perfect state transfer occurs periodically.

In this paper, we examine the system of $N$ coupled cavities and two-level atoms interacting successively with the field of their respective cavity for $A>>g$. Adjusting the detuning $\delta$ and the coupling $A$, a resonant interaction between the atom and only one delocalized normal mode of the field occurs, so we investigated the transmission of the quantum state of an atom in cavity $s$ (\textit{sender}) for an atom in cavity $r$ (\textit{receiver}). We'll show how, in the ideal case, a perfect transfer can occur independent of $s$ and $r$ (flexible) and also analyze the effects of dissipation via master equation. We also studied the presence of important sources of imperfections such as variation in the interaction time of the atoms with the cavity's field and delay between the procedure steps. 

\section{\label{sec:level2}Coupled cavities system}
\

The full Hamiltonian of the system formed by $N$ the coupled cavities and a single two-level atom interacting with the cavity field $j$ is
\begin{equation}
	H^{(j)} = H_{at}^{(j)} + H_{C} + H_{JC}^{(j)},
	\label{eq:htot}
\end{equation}
\noindent with,  
\begin{equation}
	H_{at}^{(j)}=\hbar \omega_{0} \frac{\sigma_{jz}}{2},
\end{equation} where $\omega_{0}$ the frequency of the atomic transition $\left|g_{j}\right\rangle\Leftrightarrow\left|e_{j}\right\rangle$. 

Considering a two-level atom interacting only with one cavity field with coupling constant $g$, the interaction between the atom and the field in cavity $j$ in dipole and rotating wave approximation (RWA)\cite{scully,walls} is given by the Jaynes-Cummings model (JCM)\cite{scully,walls,jcm},
\begin{equation}
H_{JC}^{(j)}=\hbar g\left(\hat{a}^{\dagger}_{j}\sigma_{j-}+\hat{a}_{j}\sigma_{j+}\right)
\label{eq:jc}
\end{equation}

The Hamiltonian that describes the photons in the coupled-cavity is \cite{angelakis,hartmann,li,yabuuti,nohama,ogden},
\begin{equation}
	H_{C}=\hbar\sum_{n=1}^{N}\omega_{c}\hat{a}_{n}^{\dagger}\hat{a}_{n} + \hbar A\sum_{n=1}^{N}\left(\hat{a}_{n}^{\dagger}\hat{a}_{n+1}+\hat{a}_{n}\hat{a}_{n+1}^{\dagger}\right)
	\label{eq:ha}
\end{equation}
\noindent where $\hat{a}_{n}(\hat{a}_{n}^{\dagger})$ is annihilation (creation) operator for the photon in cavity $n$, and $A$ is
the hopping rate of photons between neighboring cavities. For simplicity we assume a homogeneous system, ie, the cavity mode $\omega_{c}$ and the hopping rate $A$ uniform. Indeed, it is very hard to control the uniformity of the hopping rate, as it is typically fixed during
fabrication of the cavities, but the fluctuations are assumed to be innocuous and do not change the results, qualitatively. Further, in a special case, these fluctuations won't be relevant. 

The second term in \eref{eq:ha} can be derived phenomenologically \cite{yariv,haus} assuming the coupling between the cavities to be proportional
to the overlap of the fields' modes. This should be a good approximation if the cavities are not too strongly coupled ($\omega_{c}>>A$)\cite{yariv,haus}.

\begin{figure}[h]
\centering
\includegraphics[angle=-90,width=0.4\textwidth]{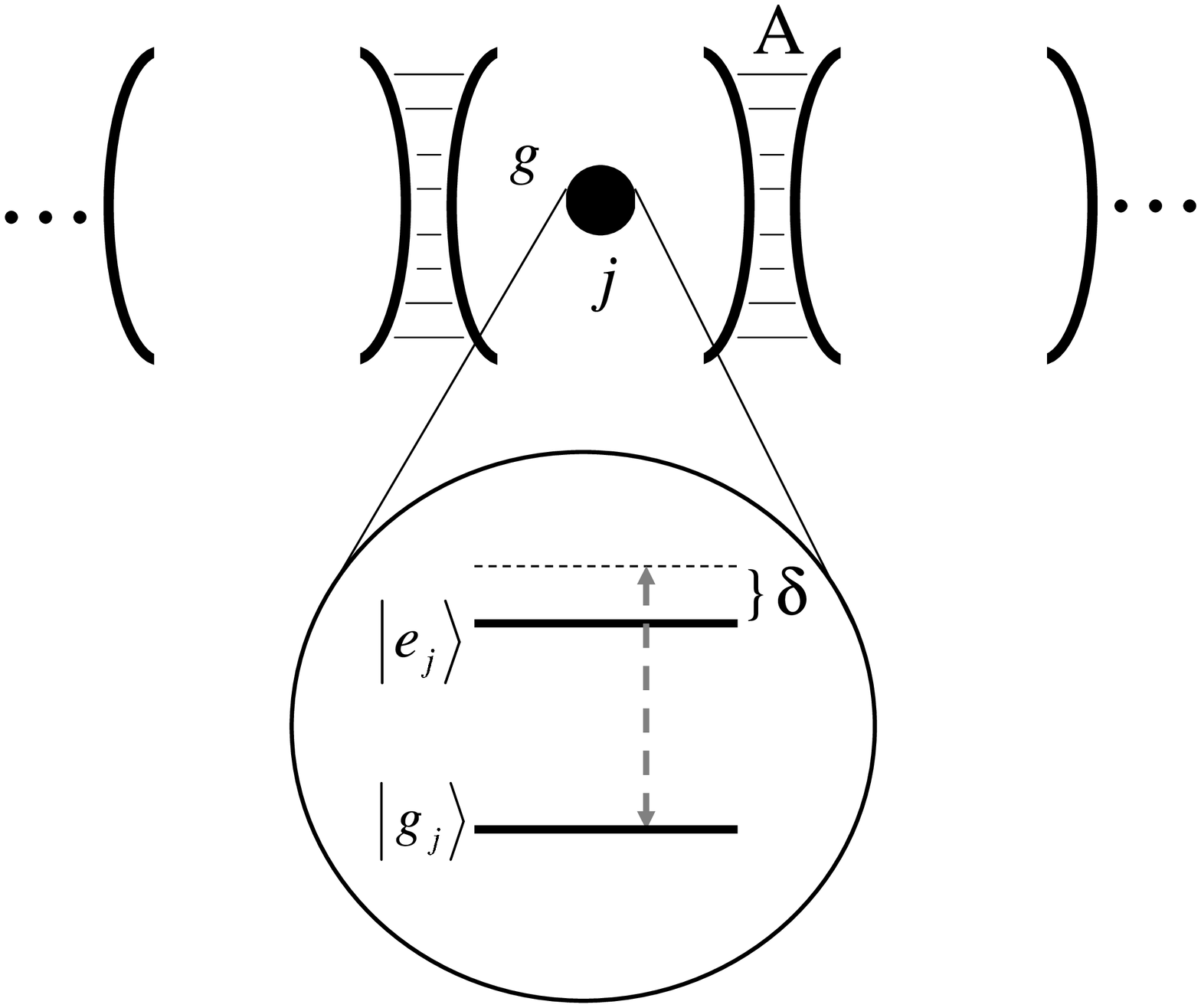}
\vspace{1cm}
\caption{Schematic diagram of a one-dimensional coupled-cavity system and one two-level atom in cavity $j$ where $\delta$ is the atom-field detuning.}
\label{fig:figure1}
\end{figure}

Let us consider the normal modes (delocalized field) of the coupled cavities system as presented in \cite{serafini,nohama,yabuuti,rai}. In such a case the Hamiltonian \eref{eq:ha} can be diagonalized using the following bosonic operators,
\begin{eqnarray}	
\hat{b}_{k}=\sum_{n=1}^{N}S\left(n,k\right)\hat{a}_{n} \\ \nonumber \hat{a}_{n}=\sum_{k=1}^{N}S^{*}\left(n,k\right)\hat{b}_{k}
\label{eq:normal}
\end{eqnarray}
\noindent that lead to,
\begin{equation}
	H_{C}=\hbar\sum_{k=1}^{N}\left(\omega_{c}+\beta_{k}\right)\hat{b}_{k}^{\dagger}\hat{b}_{k}.
\end{equation}

With appropriate boundary conditions for the system, i.e. $\hat{a}_{N+1}=0$, we have,
\begin{eqnarray}	\label{eq:beta}
	\beta_{k}=2Acos\left(\frac{k\pi}{N+1}\right), ~~~~~~~~~~~ \\ \nonumber S\left(n,k\right)=S^{*}\left(n,k\right)=\sqrt{\frac{2}{N+1}}sen\left(\frac{nk\pi}{N+1}\right).
\end{eqnarray}

In the interaction picture and in terms of the normal nodes the full Hamiltonian is, 
\begin{equation}
H_{I}^{(j)}=g\sum_{k=1}^{N}S(j,k)\left(\hat{b}_{k}^{\dagger}\sigma_{j-}e^{i(\delta + \beta_{k})t} + \hat{b}_{k}\sigma_{j+}e^{-i(\delta + \beta_{k})t}\right),
\label{eq:int}
\end{equation}
\noindent where $\delta=\omega_{c}-\omega_{0}$ is the atom-field detuning.

Although the exact dynamics of the system is rather complicated in the general case, there are regimes in which it may be simplified. Such as, setting the atomic transition equal to a normal mode frequency, i.e adjusting appropriately the detuning $\delta=-\beta_{q}$, the   Hamiltonian \eref{eq:int} becomes, 
\begin{eqnarray}
	H_{I}^{(j)}=\hbar gS(j,q)\left(\hat{b}_{q}^{\dagger}\sigma_{j-} + \hat{b}_{q}\sigma_{j+}\right)~~~~~~~~~~\\\nonumber +\hbar g\sum_{k\neq q}^{N}S(j,k)\left(\hat{b}_{k}^{\dagger}\sigma_{j-}e^{i(\beta_{k} - \beta_{q})t} + \hat{b}_{k}\sigma_{j+}e^{-i(\beta_{k} - \beta_{q})t}\right).
\end{eqnarray} 

The first part in the Hamiltonian above (that describes the transfer of energy between the resonant atom and normal mode) is independent of $t$, but in the second one (which represents the dispersive interaction between the atom and the nonresonant normal modes) it appears on the form $e^{\pm i(\beta_{k} - \beta_{q})t}$. 

From \eref{eq:beta} we know that $\beta_{k}-\beta_{q}\approx A$. So, for $A >> g$, the interaction of the atom with the nonresonant normal modes can be neglected (essentially truncating the Dyson series) and the system reduces to an atom resonantly coupled to a single-mode field \cite{serafini,lu,zhangqi,peng},
\begin{equation}
	H_{I}^{\left(j\right)}=\hbar gS(j,q)\left(\hat{b}_{q}^{\dagger}\sigma_{j-} + \hat{b}_{q}\sigma_{j+}\right).
	\label{eq:hint}
\end{equation}

In such approximation, the atom, initially interacting only with the field in cavity $j$, is coupled with the delocalized mode $q$. 
It means that the atom interacts with the field of all cavities, that has $S(n,q)\neq 0$. 

Numerical simulations were done in order to check the validity of the approximation that lead the Hamiltonian \eref{eq:htot} to \eref{eq:hint} in the limit $A>>g$ for $N=3$ \cite{serafini,lu,zhangqi,peng}. Here we assume the validity of the approach to any $N$ once the arguments are the same. 

The dynamics of the system can be determined by the evolution operator given by,
\begin{equation}\begin{array} {c}
		U^{(q)}_{j}\left(t\right)=e^{-iH_{I}^{\left(j\right)}t/\hbar}=      
        cos(S\left(j,q\right)gt\sqrt{\hat{b}_{q}\hat{b}_{q}^{\dagger}})\left|e_{j}\right\rangle\left\langle e_{j}\right|  
        \\ \nonumber+cos(S\left(j,q\right)gt\sqrt{\hat{b}_{q}^{\dagger}\hat{b}_{q}})\left|g_{j}\right\rangle\left\langle g_{j}\right| \\     -\frac{\textit{i}}{\sqrt{\hat{b}_{q}\hat{b}_{q}^{\dagger}}}sen(S\left(j,q\right)gt\sqrt{\hat{b}_{q}\hat{b}_{q}^{\dagger}}) \hat{b}_{q}\left|e_{j}\right\rangle\left\langle g_{j}\right| \\ \nonumber       -\frac{\hat{b}_{q}^{\dagger}\textit{i}}{\sqrt{\hat{b}_{q}\hat{b}_{q}^{\dagger}}}sen(S\left(j,q\right)gt\sqrt{\hat{b}_{q}\hat{b}_{q}^{\dagger}})  \left|g_{j}\right\rangle\left\langle e_{j}\right|, 
\end{array}
\end{equation}
\noindent similarly with the JCM \cite{scully,walls,jcm}.

By adjusting the detuning and keeping $A>>g$, we can simplified the dynamics of the system to a JCM. In the next section we will show how one can use such system for a perfect and flexible quantum communication. Noting that fulfilling the condition $A>>g$ might require weak couplings, thus implying larger operating times.

\section{\label{sec:level3}Quantum communication}
\

In order to analyze the quantum communication between distant atoms we will split our procedure into two steps. Initially only the atom \textit{sender}, whose state must be transmitted, interacts with its cavity's field (step $1$), and only after that, the atom \textit{receiver} goes into action (step $2$). The two atoms never interact simultaneously with the coupled cavities system in this procedure.

It is necessary a minimal control, that is, only the sender and the receiver can apply local  operations to the system and the other parts of the array cannot be controlled during the communication process. A protocol with less control is obviously impossible \cite{burgarth}. 

Control the atom-field interaction in a particular cavity per time can be done with high accuracy during the atomic passage via
Stark shift tuning (i.e., the atomic levels can be shifted through a static electric field applied in the cavity) or just by selecting the appropriate atomic velocity. Nowaday, this can be done with Rydberg atoms and microwave cavities \cite{raimond}. On the other hand, strong coupling between transported atoms and optical microcavities with open access had been observed recently \cite{sauer,aoki,trupke,barclay}.

In the first step of our procedure we assume that the coupled cavities system is prepared in its vacuum state, the atom $r$ in the ground state and the atom $s$ in an arbitrary state to be transmitted, i.e,
\begin{equation}	\left|\psi_{i}\right\rangle=\left(cos(\frac{\theta}{2})\left|g_{s}\right\rangle+e^{i\phi}sen(\frac{\theta}{2})\left|e_{s}\right\rangle\right)\otimes\left|g_{r}\right\rangle\otimes\left|0\right\rangle.
\end{equation} 

 Then, the fidelity of the quantum communication through the array for a arbitrary pure state, as a function of the interaction time $t_{1}$ (step $1$) and $t_{2}$ (step $2$), is 
\begin{eqnarray}
	F_{s,r}^{(q)}\left(\theta,\phi,t_{1},t_{2}\right)=\left|\left\langle \psi_{f}\right|U^{(q)}_{r}\left(t_{2}\right)U^{(q)}_{s}\left(t_{1}\right)\left|\psi_{i}\right\rangle\right|^{2}=\\  \nonumber cos^{4}\left(\frac{\theta}{2}\right) + sen^{4}\left(\frac{\theta}{2}\right)f_{s,r}^{(q)}\left(t_{1},t_{2}\right)^{2} \\+ \nonumber 2sen^{2}\left(\frac{\theta}{2}\right)cos^{2}\left(\frac{\theta}{2}\right)f_{s,r}^{(q)}\left(t_{1},t_{2}\right), \label{eq:Fsr} 
\end{eqnarray}
where $\left|\psi_{f}\right\rangle=\left|g_{s}\right\rangle\otimes\left(cos(\frac{\theta}{2})\left|g_{r}\right\rangle+e^{i\phi}sen(\frac{\theta}{2})\left|e_{r}\right\rangle\right)\otimes\left|0\right\rangle$ and,
\begin{equation}
	f_{s,r}^{(q)}\left(t_{1},t_{2}\right)=-sen(S\left(s,q\right)gt_{1})sen(S\left(r,q\right)gt_{2}).\label{eq:fsr}
\end{equation}

This function, that measures the quality of the transfer, is always between $0$ and $1$. Higher value meaning better transmission and $F_{s,r}^{(q)}=1$ is a perfect transfer. Fidelity of $2/3$ can be obtained just using local operations and classical communication. Thus, $F_{s,r}^{(q)}$ needs to be greater than $2/3$ in the quantum communication protocol to be better than a pure classical channel.

In general, for $S(s,q)gt_{1}=\pm\pi/2$, the exciton that is initially in atom $s$, is transferred to the normal mode $\hat{b}_{q}$,
\begin{equation}	
\left|g_{s}\right\rangle\otimes\left|g_{r}\right\rangle\otimes\left(cos(\frac{\theta}{2})\left|0\right\rangle \mp ie^{i\phi}sen(\frac{\theta}{2})\left|1\right\rangle\right),
\end{equation}
\noindent and in second step, at $S(r,q)gt_{2}=\pm 3\pi/2$, the exciton goes to the atom $r$,
\begin{equation}	
\left|g_{s}\right\rangle\otimes\left(cos(\frac{\theta}{2})\left|g_{r}\right\rangle+e^{i\phi}sen(\frac{\theta}{2})\left|e_{r}\right\rangle\right)\otimes\left|0\right\rangle,
\end{equation}
\noindent so substituting these specific times in \eref{eq:fsr} and \eref{eq:fsr} we obtain, in the ideal case, $F_{s,r}^{(q)}=1$, i.e., a perfect transfer for any $N$, $s$ (\textit{sender}) and $r$ (\textit{receiver}) such that $S(s,q)\neq 0$ and $S(r,q)\neq 0$, independently of the initial state.

Apart of the perfect transmition of quantum states for any $N$, in the case of $N$ odd the normal mode $q=(N+1)/2$ give us $\beta_{q}=0$ and,
\begin{center}
\begin{math}
	S\left(j,q\right)= \left\{\begin{array}{c}  
       0~~~~~~~~~j~even\\   
       \\    
       \left(-1\right)^{(j-1)/2}\sqrt{\frac{2}{N+1}}~~~~~j~odd\\ 
\end{array}\right.
\end{math}
\end{center}

Although, according to the Hamiltonian \eref{eq:hint} there is no interaction between crossing atoms and even cavities, i.e, we miss the capacity of realize the quantum communication through the system via even cavities, we gain the simplification $\delta=0$. Otherwise we would have to adjust the detuning depending on the coupling constant $A$, an experimental difficulty. 
\begin{table}[h!!]
	\caption{\label{tab:srq}Interaction time $gt_{1}$ and $gt_{2}$ that result $F_{s,r}^{(q)}=1$ with $N=3,5$ and $7$.}
\begin{indented}
\lineup
\item[]\begin{tabular}{@{}*{4}{l}} 
\br 
        & $s~\rightarrow~r$   &  $gt_{1}$ & $gt_{2}$ \cr  
\br         
   $N=3,~q=2~~~~$& $1~\rightarrow~3$   &   $\pi/\sqrt{2}$    &    $\pi/\sqrt{2}$     \cr  
\mr 
        & $1~\rightarrow~3$   &   $\sqrt{3}\pi/2$    &    $\sqrt{3}\pi/2$      \cr 
   $N=5,~q=3~~~~$& $1~\rightarrow~5$   &    $\sqrt{3}\pi/2$   &     $3\sqrt{3}\pi/2$    \cr  
        & $3~\rightarrow~5$   &   $\sqrt{3}\pi/2$   &   $\sqrt{3}\pi/2$     \cr  
\mr         
        & $1~\rightarrow~3$   &   $\pi$    &    $\pi$      \cr 
        & $1~\rightarrow~5$   &    $\pi$   &     $3\pi$    \cr  
   $N=7,~q=4~~~~$& $1~\rightarrow~7$   &   $\pi$   &   $\pi$     \cr   
        & $3~\rightarrow~5$   &   $\pi$   &   $\pi$     \cr  
        & $3~\rightarrow~7$   &   $\pi$   &   $3\pi$     \cr 
        & $5~\rightarrow~7$   &   $\pi$   &   $\pi$    \cr  
\br          			
\end{tabular}
\end{indented}
\end{table}

Furthermore, in this case, the fluctuations  during the process of fabrication of the array resulting in  various values of the constant $A$ is no more relevant. We don't need to know specifically the value of the hopping rate just only to have $\omega_{c}>>A>>g$.

\section{\label{sec:level4}Master Equation: dissipation effects}
\

In this section we will study the efficiency of the quantum state transfer under dissipation. A realistic quantum system can not be isolated from its environment completely, rather it is usually coupled to a large number of degrees of freedom called heat bath. There are two main loss processes which are serious obstacles against the goal of an efficient quantum communication: spontaneous emission from the excited state to the ground state due to its interaction with the modes outside the cavities, and leaking out of photons. The latter is even worse with the increase of $N$. 

As, in the beginning, we assume all the cavities in the vacuum state and in \eref{eq:hint} the nonresonant normal are not involved, they will never be populated and can be neglected in the master equation. Considering also the weak coupling $\left(\omega_{c}>>A\right)$, the indirect channels of decay vanish \cite{ponte}.

Based on such considerations and in the Born-Markov approximation (considering the temperature of the reservoir $T=0K$), the master equation for the density matrix of the system  in step $1$ may be written as
\begin{eqnarray}	\frac{d}{dt}\rho=-\frac{i}{\hbar}\left[H_{I}^{(s)},\rho\right] \nonumber+\frac{N\gamma}{2}\left(2\hat{b}_{q}\rho\hat{b}_{q}^{\dagger}-\hat{b}_{q}^{\dagger}\hat{b}_{q}\rho-\rho\hat{b}_{q}^{\dagger}\hat{b}_{q}\right) \\+\frac{\kappa_{s}}{2}\left(2\sigma_{s-}\rho\sigma_{s+}-\sigma_{s+}\sigma_{s-}\rho-\rho\sigma_{s+}\sigma_{s-}\right),
\end{eqnarray}
\noindent where $\kappa$ denote the spontaneous decay rate of the excited atom state, $\gamma$ is the decay rate of cavity field (equal for all the cavities) and $\tilde{H}^{(j)}$ is the Hamiltonian \eref{eq:hint} with $j=s$ (step $1$). 

After the first atom interact with its cavity field we started the second step, in which the second atom interacts with the field in cavity $r$ (remembering that we should still consider the effects of dissipation in the first atom),
\begin{eqnarray}	\frac{d}{dt}\rho=-\frac{i}{\hbar}\left[H_{I}^{(r)},\rho\right]  \nonumber+\frac{N\gamma}{2}\left(2\hat{b}_{q}\rho\hat{b}_{q}^{\dagger}-\hat{b}_{q}^{\dagger}\hat{b}_{q}\rho-\rho\hat{b}_{q}^{\dagger}\hat{b}_{q}\right)\\+\sum_{j=s,r}\frac{\kappa_{j}}{2}\left(2\sigma_{j-}\rho\sigma_{j+}-\sigma_{j+}\sigma_{j-}\rho-\rho\sigma_{j+}\sigma_{j-}\right).
\end{eqnarray}
 
The efficiency of the state transfer will be given by the mean value of the fidelity of quantum communication over all initial states in Bloch sphere, 
\begin{equation}
	\bar{F}_{s,r}^{(q)}=\left\langle F_{s,r}^{(q)}\right\rangle_{\theta,\phi},
\end{equation}
\noindent with $F_{s,r}^{(q)}=\left\langle \psi_{f}\right|\rho\left|\psi_{f}\right\rangle$. The maximum average fidelity is achieved at time when the quantum state transfer is supposed to happen perfectly.

\begin{table}[h!!]
\caption{\label{tab:srqdiss}Maximum average fidelity achieved for $N=3,5,7$.}
\begin{indented}
\lineup
\item[]\begin{tabular}{@{}*{3}{l}} 
\br 
                 & $s~\rightarrow~r$   &  $~~~\bar{F}_{s,r}^{(q)}$ \cr 
\br                 
   $N=3,~q=2~~~~$& $1~\rightarrow~3$   &   $~~~0.986$        \cr  
\mr    
                 & $1~\rightarrow~3$   &   $~~~0.977$         \cr 
   $N=5,~q=3~~~~$& $1~\rightarrow~5$   &   $~~~0.955$      \cr  
                 & $3~\rightarrow~5$   &   $~~~0.977$        \cr  
\mr                  
                 & $1~\rightarrow~3$   &   $~~~0.965$         \cr 
                 & $1~\rightarrow~5$   &   $~~~0.934$      \cr  
   $N=7,~q=4~~~~$& $1~\rightarrow~7$   &   $~~~0.965$       \cr   
                 & $3~\rightarrow~5$   &   $~~~0.965$      \cr  
                 & $3~\rightarrow~7$   &   $~~~0.934$       \cr  
                 & $5~\rightarrow~7$   &   $~~~0.965$      \cr 
\br                    				
\end{tabular}
\end{indented}
\end{table}

We can observed in \tref{tab:srqdiss} that the maximum average fidelity presented values above $0.93$ for $N=3,5,7$ and $q=(N+1)/2$ with currently realizable experimental parameters $\gamma=0.004g$ and $\kappa_{s}=\kappa_{r}=0.006g$\footnote{Here we will consider $\kappa_{s}=\kappa_{r}$ (identical atoms in the same environment), but in fact, if we consider passage atoms the rate of spontaneous emission depends on the environment in which the atoms are inserted (Purcell effect \cite{purcell}).} (assuming $g=2\pi\times 450$ MHz) involving the interaction between cesium atoms and optical microtoroidal cavities at resonance with the $D_{2}$ transition of cesium which occurs at a wavelength of $852.359 nm$\cite{spillane}.

For instance, for $N=7$, starting in cavity $s=1$ one can transfer an arbitrary quantum state to $r=3,5,7$ with a good fidelity, indicating that
the protocol is flexible under these conditions. Moreover, the starting point can be any cavity and the process cen be reversed, i.e, the state return to the original cavity.

\begin{figure}[h!]
\centering
\includegraphics[angle=-90,width=0.4\textwidth]{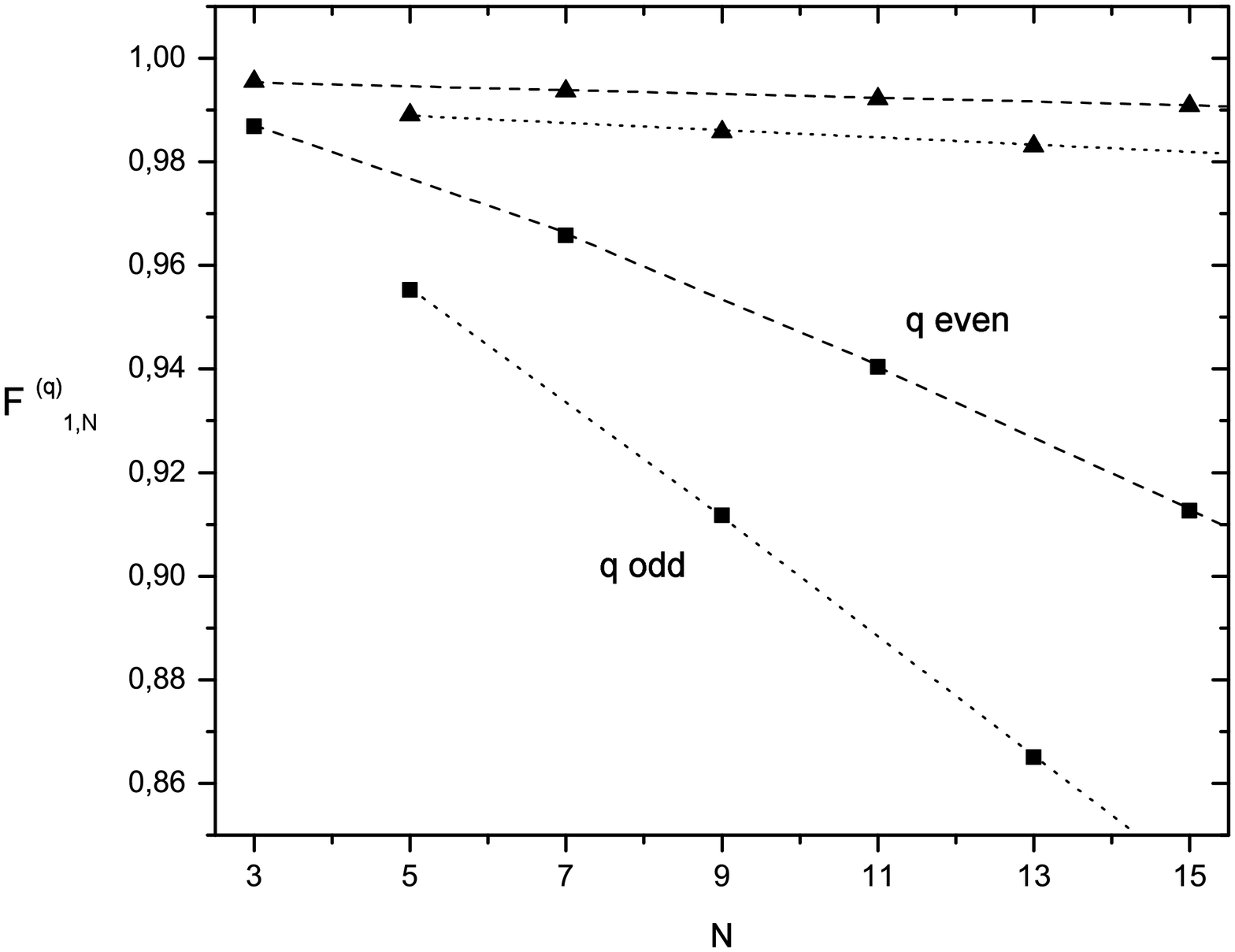}\vspace{1cm}
\caption{Maximum average fidelity as a function of the total number of cavities $N$: (square) current experimental parameters, (triangle) optimizing the strong coupling in the projected limit for microtoroidal optical cavities\cite{spillane}.}
\label{fig:figure2}
\end{figure}

It is interesting to note a peculiar parity effects for the normal mode $q$ selected to interact resonantly with the two-level atom. The fidelity is higher for even $q$ rather than the odd ones, for example, since $q=(N+1)/2$,  $N=7$ conduces to $q=4$, with result $\bar{F}_{1,7}^{(4)}=0.965$ while for $N=5$ ($q=3$)we have $\bar{F}_{1,5}^{(3)}=0.955$, a result obviously not trivial. 
The reason is a different phase in $S(N,q)$ for even-odd $q$,
\begin{equation}
		S\left(N,q\right)=\left(-1\right)^{q-1}\frac{1}{\sqrt{q}},
\end{equation}
\noindent for even $q$ is needed a smaller time to transfer the state, then the damage caused by the dissipation is less than for a odd $q$.
In \fref{fig:figure2} such peculiar behavior in the parity is best observed, dashed line is for $q$ even and dot line for $q$ odd. 

Besides the flexibility, another important feature for a quantum communication protocol is its extension. So, we now want to investigate the performance of our protocol for various array lengths $N$ with $s=1$ and $r=N$ under dissipation. The maximum average fidelities as a function of $N$ are shown in \fref{fig:figure2} assuming only odd $N$ and with $q=(N+1)/2$ ($\delta=0$).

With the increase of $N$ the maximum fidelity decreases as the total time required for communication is greater ($gt\propto\sqrt{N}$). Arrays of length greater than $51$ have fidelity smaller than $2/3$ (this is the fidelity of the best classical transmission of an unknown qubit \cite{horodecki}), thus the procedure is no longer useful for $N>51$ considering current experimental parameters. On the other hand, assuming the projected limit for microtoroidal optical cavities and cesium atoms optimizing the strong coupling regime with cooperativity factor $g^{2}/\gamma\kappa_{j}\approx 10^{7}$ is possible to obtain a fidelity of $0.9$ with a array of $N=199$.

\newpage
\section{\label{sec:level5}Error in procedure execution}
\

The procedure presented in section \eref{sec:level3} require that two atoms interact, during a specific time, successively with its cavity mode. In realistic setups, however, it might happen a slight variation in the interaction time or a delay between the two steps of the procedure can occur, consequently inducing an error in the communication. Here we will not consider an error due a simultaneously interaction, in such case the dynamic changes completely. 

\begin{figure*}[h]
\centering\vspace{1.5cm}
\includegraphics[width=0.35\textwidth]{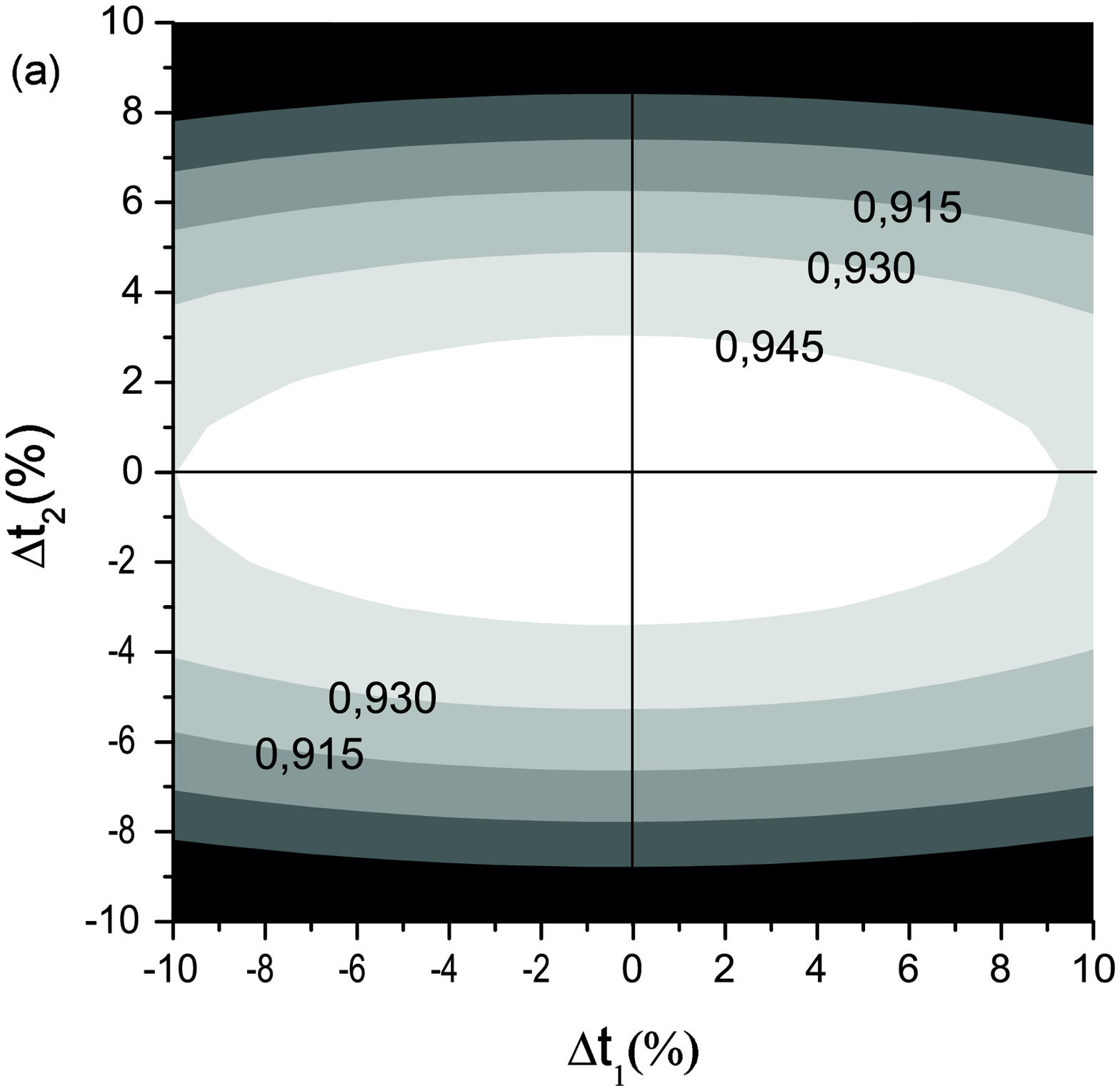}\hspace{1.5cm}
\includegraphics[width=0.35\textwidth]{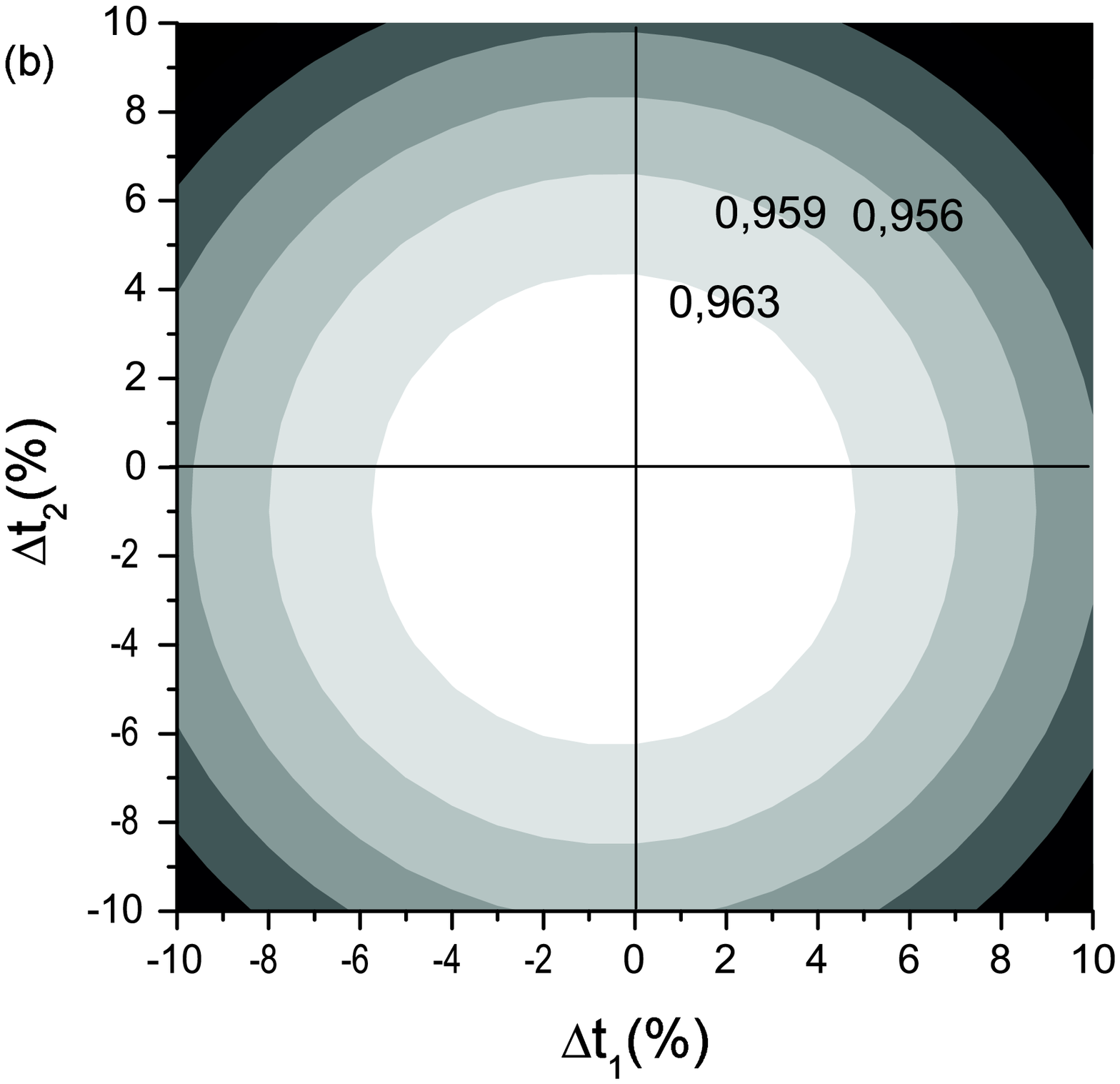}\vspace{1.5cm}
\includegraphics[width=0.35\textwidth]{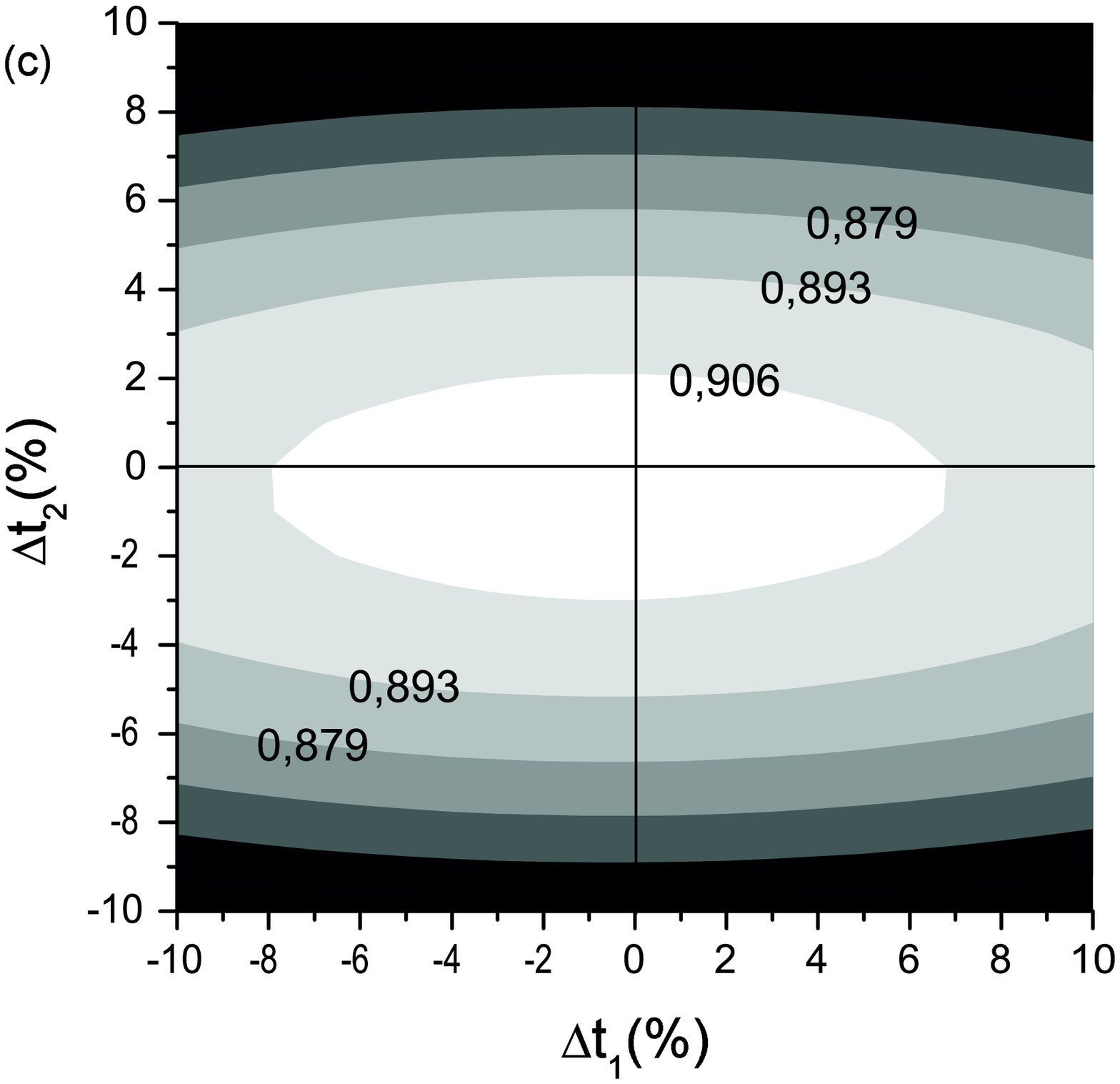}\hspace{1.5cm}
\includegraphics[width=0.35\textwidth]{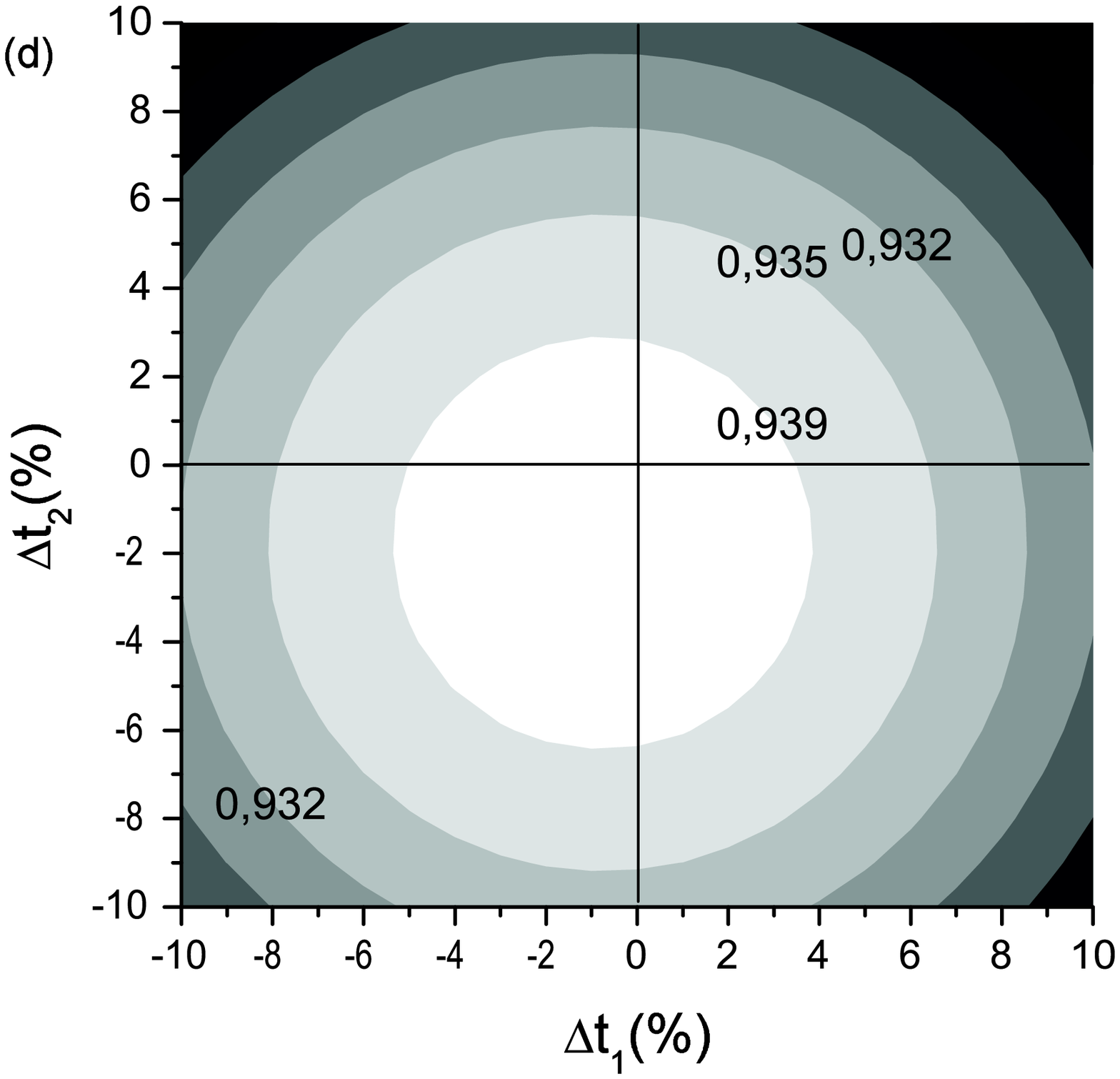}
\caption{Fluctuation in the average fidelity as function of the variation in the interaction time $\Delta t_{1}$ and $\Delta t_{2}$ for the quantum communication between $s=1$ and $r=N$ with $q=(N+1)/2$ and $N=$(a)$~5$,(b)$~7$,(c)$~9$,(d)$~11$ considering current experimental parameters \cite{spillane}.}
\label{fig:figure3}
\end{figure*}

In \fref{fig:figure3} we show how the average fidelity of quantum communication between the ends of a array with $N=5,7,9$ and $11$ behaves assuming a percentage variation in the interaction time, the cross line in the middle of the graphic represents the ideal interaction time. We can note that a positive percentage error ($\Delta t_{i}>0$, $i=1,2$) does more harm to the fidelity since it gives more time to dissipate but even with a error of $2\%$ in both steps of the procedure the fidelity of transmission keeps above $0.9$.

The damage effects into $F$ due errors in the interaction time can be minimized if one measure the atom $s$ after the step $1$ although this affects the sucess probability (without a measurement the protocol is deterministic). If the atom is obtained in the excited state, the protocol obviously failed and should be repeated. On the other hand, if $g$ is obtained, the exciton was transferred to the field and the procedure can be completed with a certain fidelity (better compared with the case without measurement) and a probability of success. 

\begin{figure}[h]
\centering
\includegraphics[angle=-90,width=0.4\textwidth]{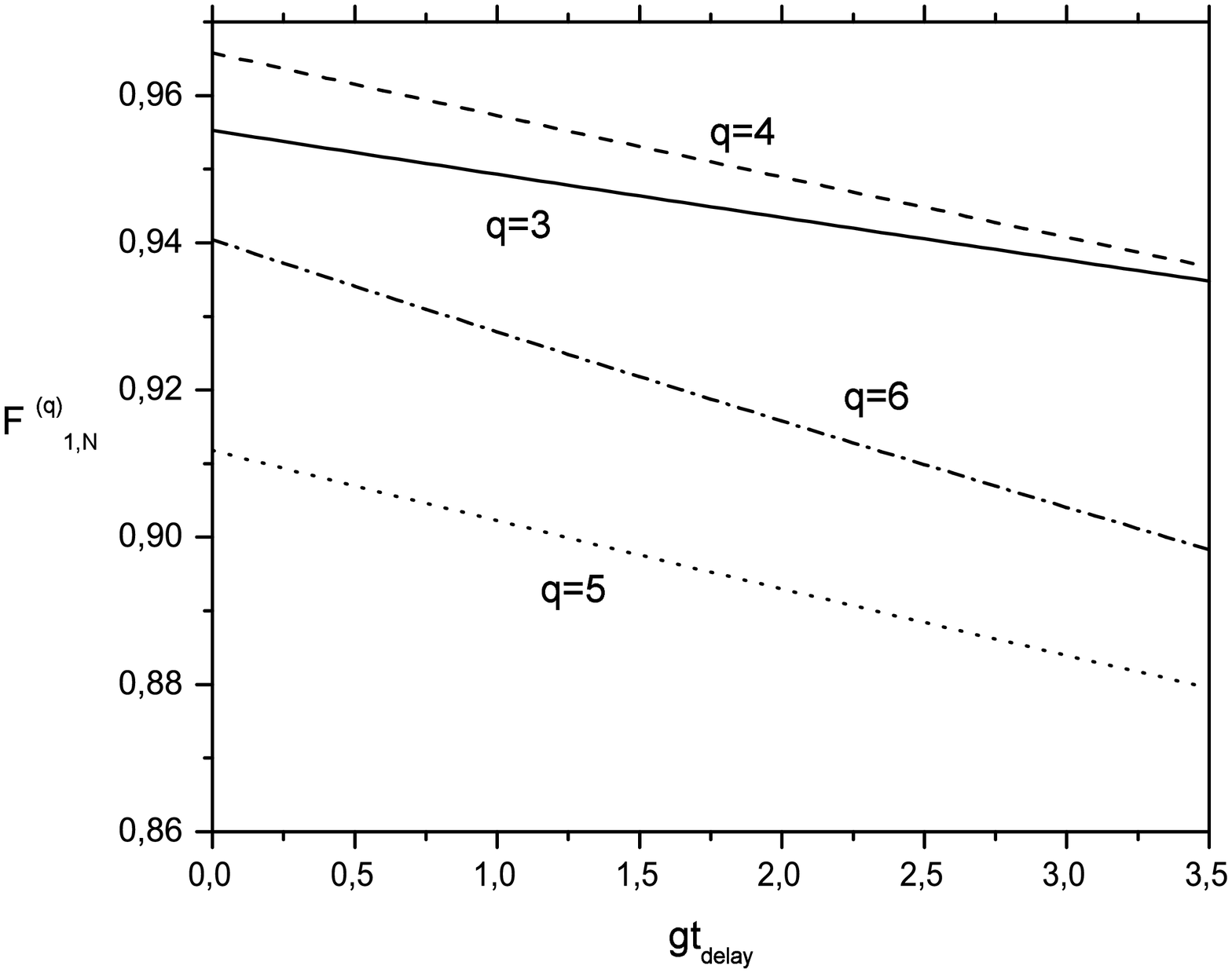}\vspace{1cm}
\caption{Average fidelity as function of a delay time between the two steps of the procedure for a quantum transfer between the ends of the array for $N=$(solid)$~5$,(dash)$~7$,(dot)$~9$,(dash dot)$~11$.}
\label{fig:figure4}
\end{figure}

Now, if the second step don't start exactly after the first, the fidelity of the quantum state transfer is damaged as we can see in \fref{fig:figure4} for $N=5,7,9,11$ again with $q=(N+1)/2$ and using current experimental parameters \cite{spillane}. Although the fidelity with $N=7$ (q even) is larger than $N=5$ (q odd) with $gt_{delay}=0$, the delay in the procedure affects more the system with seven cavities ($N=7$) than five cavities, since there are more cavities involved in dissipation channel.  After a time around $gt_{delay}=4.5$ the lines (solid $N=5$ and dash for $N=7$) intersect. For a long delay the fidelity tends to $1/2$, however it is important to see that a delay of approximately $gt_{delay}\approx \sqrt{N}$ give us a fidelity still above $0.88$ for $N=5,7,9$ and $11$.

\section{\label{sec:level6}Discussion}
\

The dynamics of the system is governed by three basic parameters: $g$, $\delta$, and $A$. Two limits are essential for the model, $\omega_{c}>>A$ and $A>>g$. The first one intended to validate the Hamiltonian of the coupled cavity system and the second one to eliminate the nonresonant normal modes. Fixing $\delta=-\beta_{q}$, the atom interacts resonantly with the normal mode $\hat{b}_{q}$ and the interaction of the atoms with the nonresonant normal modes can be neglected summarizing the system to Jaynes-Cummings model.

The procedure for quantum state transfer described here can be performed on a system of atoms interacting with a single cavity. However, the coupled cavity system offers some advantages. Distinct cavities can be more easily isolated from each other while avoiding cross-talk. Moreover, the separation allows an individual access to each cavity, controlling and measuring the atom-field systems separately.

In summary, we propose a simple but robust quantum communication procedure between distant atoms in a system of $N$ coupled cavities. We achieved a perfect state transfer for any number of cavities $N$ in the ideal case for $A>>g$. In fact, the procedure is also flexible, i.e. independent of the location of the \textit{sender} and the \textit{receiver} in the CCA.

It is also important in our protocol that after finishing the communication, the CCA is again in the vacuum state and another state may be transmitted without any initial preparation.

For a odd number $N$ of coupled cavities with $\delta=0$ (experimental simplification) we selected the normal mode $q=(N+1)/2$ to interact resonantly with the two-level atom. In such a scenario we study two sources of imperfections: the effects of dissipation by spontaneous decay and cavities losses; and errors in the procedure execution. 

We obtain that, in spite of these conditions, the average fidelity is still able to reach values which are approximately equal to the unit for $N=3,5$ and $7$, besides that, until $N<51$ the protocol is more effective than any classical channel showing that the procedure for obtaining the quantum state transfer is reliable. Futhermore, even under imperfections during the process the results showed good fidelities for subtle variations of the interaction time and delay between the steps.

The novel even-odd parity effect presented here (in our view, for the first time in the hybrid system of atom and coupled cavity array) should be observed experimentally and it is a good evidence for the validity of the model.

There are promising candidates for experimental implementation of the quantum communication protocol in the hybrid system of $N$ coupled cavity with atoms.   

One important candidate consists of quantum dots coupled to $2D$ photonic crystals nanocavities with high-Q factors and small mode volumes \cite{akahane,altug,na}. Photonic crystals are periodic dielectric structures with properties that affect the motion of photons, analogous to a semiconductor whose periodicity influences the movement of electrons. In this scenario, nanocavities are localized defects in the crystal structure so that the radiation with a particular frequency is confined in the defect.

Quantum dots (which can be considered a two-level system similar to atoms in the JCM \cite{jcm}) can be created within these nanocavities \cite{badolato} and interact with the field mode forming a standard CQED system. Moreover, due to its small mode volume, the atom-field coupling $g$ can be extremely large \cite{akahane} while a strong coupling regime had already been obtained experimentally \cite{hennessy}.

The most promissing experimental implementation (in our view), which can be produced in an array with high-$Q$, are disk \cite{barclay} or toroidal \cite{spillane} silicon microstructures.

A near lossless coupling between these microstructures is possible via fiber taper placed close to their surface,
evanescent waves of cavity and fiber overlap and photons can tunnel between them \cite{barclay,spillane.043902}. The coupling can be controlled by the separation between the cavity and fiber.
Moreover, the field in such structures can interact with cesium atoms placed close to the cavity surface in a strong coupling regime, with a very large cooperativity factors \cite{barclay,spillane}.


In summary, setting the distance between the microstructures during the manufacturing, such that $A\approx 50GHz$, we have $\omega_{c}>>A>>g$ (for $g\propto GHz$) and hence the model presented here may be feasible in the current state of technology.


Working with a set of $M$ atoms (instead of only one), Zhang et. al. \cite{zhangqi} obtained a speed-up in the operation time of the quantum state transfer which is proportional to $1/\sqrt{M}$. This is useful for depressing the influence of unavoidable decoherence processes such as spontaneous emission and photon leakage. In our procedure, such a speed-up can be useful too.

\ack
B.F.C.Y thanks the financial support from Conselho Nacional de Desenvolvimento Cient\'ifico e Tecnol\'ogico (CNPQ). J.A.R. thanks CNPq for partial support of this work. J.A.R. also acknowledge partial support of CEPOF (Centro de Pesquisa em \'Optica e Fot\^onica).

\section*{References}
\bibliographystyle{iopart-num}
\bibliography{myrefs}

\end{document}